\documentstyle[aps]{revtex}
\begin{document}
%.
\draft
\title{
Universal Level Dynamics of Complex Systems}
\author{Pragya Shukla$^{*}$} 
\address{Department of Physics,   
Indian Institute of Science, Banglore-560012, India.}

\maketitle
\begin{abstract}
	We study the evolution of the distribution of eigenvalues of a 
$N\times N$ matrix subject to a random perturbation drawn from  
(i) a generalized Gaussian ensemble (ii) a non-Gaussian ensemble 
with a measure variable under the change of basis. It turns out that,
in the case (i), a redefiniton of the parameter
governing the evolution leads to a Fokker-Planck equation similar 
to the one obtained when the perturbation is taken from a standard
Gaussian ensemble (with invaraiant measure). This  equivalence  can
therefore help us to obtain the correlations for various
physically-significant cases modeled by generalized Gaussian ensembles
by using the already known correlations for standard Gaussian ensembles.  

For large $N$-values, our results for both cases (i) and (ii)  
are similar to those obtained for Wigner-Dyson gas as well as for the
perturbation taken from a standard Gaussian ensemble. This seems to suggest
the independence of evolution, in thermodynamic limit, from the nature of 
perturbation involved  as well as the initial conditions and therefore
universality of  dynamics of the eigenvalues of complex systems.

\end{abstract}
\section{Introduction}
\label{int}
.

	Many important physical properties of complex quantum systems 
can be studied by analyzing the statistical response of eigenvalues to 
an external perturbation. It turns out that the evolution of the 
eigenvalues of the system as a function of the strength of the 
perturbation is universal but only after an appropriate normalization 
of the perturbing potential [1,2]. Various approches have beeen adopted 
for a better understanding of this phenomena e.g. non-linear $\sigma$ 
model [3], random matrix theory (RMT) [4], semi-classical techniques [5]etc. 
Among these, 
a special class of Gaussian ensembles of RMT have been used very successfully  
in modelling the short energy-range behaviour of eigenvalues [4]. The probability 
density of a matrix in this ensemble depends only on the functions invariant under 
change of basis e.g trace of the matrix thus making it easier to calculate 
the eigenvalue distribution.  
However it has been conjectured that the local 
statistical properties of a few eigenvalues of a random hermitian 
matrix are independent of the details of the distribution of the 
matrix elements apart from 
a few broad characterizations such as real symmetric versus complex 
hermitian etc [4].
Therefore the success of RMT in modelling the energy-level behaviour of 
complex systems should not be restricted only to the standard Gaussian 
ensembles. 

          In this analytical work, we study this conjecture by 
examining the 
dynamics of eigenvalues of a quantum hamiltonian under a random perturbation 
belonging to a class of (i) a general Gaussian ensembles and (ii) a
 non-Gaussian ensemble with logarithm of the distribution given by a polynomial. 
In both cases, the 
probability density of a matrix is chosen to contain the  basis-dependent functions  
and therefore is no longer invariant under the basis-transformation. 
By integrating over all perturbations from such an ensemble, we 
show that, for large $N$, the probability density of eigenvalues evolves according 
to a Fokker-Planck (F-P) equation similar to the one proposed by Dyson for
Wigner-Dyson gas [6]. The strength 
of the perturbation acts here as a time like coordinate.  
For finite $N$, this analogy seems to survive only when the dynamics 
is considered as a 
function of the perturbation strength as well as the variances of the 
matrix elements of the perturbation.

	The statistical response of eigenvalues to a changing perturbation 
was studied by Simons and coworkers [1] by using a different approach, namely, 
non-linear $\sigma$ model.
By considering a hamiltonian 
$H=H_0+xV$ with $V$ as the perturbation and $x$ its strength, they explicitly 
calculate the  
 autocorrelation function of the energy eigenvalues for disordered 
systems within the so called Gaussian approximation (or zero mode 
aproximation). Other results pertain 
to the numerical study of 
perturbed quantum chaotic systems, such as the chaotic billiard with a
varying magnetic flux [7]. As shown in their work, the universality manifests 
itself among the 
density correlators only after the normalization of eigenvalues $\lambda_i$  
 by mean 
level spacing and that of perturbation by rms "velocity" of eigenvalues
$<(\partial\lambda_i/\partial x)^2>$; by universality one means here the independence 
of the bulk correlators in a many body system from boundary effects.   They also conjectured
 the asymptotic correspondence of the two-variable parametric density correlators 
 of complex quantum systems with the 
corresponding time-dependent particle-density correlators for the Sutherland 
hamiltonian [8].
 The connection between the static correlators (for a given parameter value) and the 
ground state correlators of Sutherland is already well known [1].
	
This correspondence was analytically shown in a recent work 
 by Narayan and Shastry [9]. They studied  
 the evolution 
of the distribution of eigenvalues of a $N \times N$ matrix $H$ subject 
to a random perturbing matrix $V$ taken from a standard Gaussian (SG) ensemble 
and showed that it leads to the F-P equation postulated 
by Dyson. Within this model, they proved the equivalence between the space-time 
correlators 
of the ground state dynamics of an integrable 1-D interacting many body quantum 
system (the Calogero-Sutherland system) and the second order parametric 
correlations of $H$.
This analogy follows because the Dyson's F-P equation is equivalent, under a Wick rotation, to the 
quantum mechanics of Calogero model which, in thermodynamic limit, has identical 
bulk properties as the Sutherland system [9,10]. 

	Our present work extends this analogy further 
 to a more general model of hamiltonians $H$. We proceed as follows. 
 For the clarification of 
our ideas, we first study, in section II, the distribution of eigenvalues 
$P(\mu,\tau)$ for an arbitrary initial condition $H_0$ and the perturbation 
$V$ taken from a generalized Gaussian 
ensemble (later referred as GG case). We obtain a partial differential equation 
governing its evolution 
which, after certain parametric redefinitions, turns out to be formally the 
same as the Fokker-Planck equation govening the evolution of Wigner-Dyson (WD) gas.
As the solution of F-P equation as well as various correlatios in the latter 
case are already known for many initial condiitions, this helps us to obtain  
 $P(\mu,\tau)$ as well as the correlations  in the GG case, as given in section III. 
 The section IV deals with the evolution of eigenvalues under a more general 
perturbation. We are motivated to study this case because 
the claim about universality of the dynamics can only be made if 
the equivalence is proved, at least in thermodynamic limit, for a $V$ with 
distribution of more general nature. We conclude in section V which is 
followed by an  appendix containing the proofs of some of the results 
used in this paper.

\section{Generalized Gaussian Case} 

Our interest is in the evolution of eigenvalues of 
$H$ under a random perturbation $V$ with initial state $H_0$ as in Ref.[9]. 
Here both $H_0$ and 
$V$ are hermitian and can be assumed to have same mean spacing without 
any loss of generality. To ensure the same mean spacing for $H$ also, we need to 
modify the parameterization of our perturbation strength $x$ [9,10,11] from $H(x)=H_0+xV$ 
to   
$H(x) = H_0 Cos(\Omega x) + V\Omega^{-1} Sin(\Omega x)$ 
with $\Omega \propto {1\over \sqrt N}$.  
Since $\Omega \rightarrow 0$ for large $N$, 
the reparameterization of the  
perturbation strength is inconsequential for finite $x$. The distribution $\rho (V)$ 
of matrix $V$ is still chosen to be a Gaussian,  
 $\rho (V)=C{\rm exp}({-\sum_{k\leq l}
\alpha_{kl} V_{kl}^2 })$ with $C$ as the normalization constant. 
So far we are in the same situation as in Ref.[9] but now we depart and 
allow the variances for diagonal and  off-diagonal 
matrix elements to be arbitrary (later referred as GG case) unlike the 
SG case where variance of the diagonal ones was twice of the off-diagonal 
ones.

        As indicated in [9], the evolution of eigenvalues of $H$ as a function 
of $x$ for any perturbation $V$ can be expressed as a set of first order differential 
equations. They turn out to be analogous to the equations of motion of a classically 
integrable system 
with $2N+\beta N(N-1)/2$ degrees of freedom (with $\beta$ a symmetry dependent 
parameter). In principle, the distribution of 
eigenvalues can be obtained from these equations and its evolution can 
be studied but the coupling between the equations makes such a method 
very difficult.  Thus we adopt another approach, the one used in [10] for studying 
the symmetric Gaussian case. As shown in [9,10], the proper form of F-P equation can be 
achieved only after a further reparametrization, namely,          
$\tau=-\Omega^{-2} {\rm ln Cos}(\Omega x)$ and by studying the evolution  
in terms of $\tau$. This suggests us to use the same parametrization for our 
case also which gives 

\begin{eqnarray}
H(\tau) = H_0 fh + V h\Omega^{-1}  
\end{eqnarray}
with $h=\sqrt{1-{\rm e}^{-2\Omega^2\tau}}$ and $fh={\rm e}^{-\Omega^2 \tau}$.

Given an ensemble of the matrices $V$ distributed with a 
probability density $\rho(V)$ in the Hermitian matrix space, 
let $P({\mu},\tau)$ be the probability of finding eigenvalues 
$\lambda_i [V]$  of $H$ between $\mu_i$ and $\mu_i+d\mu_i$ at 
"time" $\tau$ for an arbitrarily chosen $H_0$ 
which can be expressed as follows,
\begin{eqnarray}
P({\mu_i},\tau})= \int
{\prod_{i=1}^{N}\delta(\mu_i-\lambda_i[V]) \rho (V){\rm d}V 
\end{eqnarray}
Note here that $P(\mu,\tau)$ is in fact the conditional probaility 
$P(\mu,\tau|\mu_0)$ with $\mu_0$ as the eigenvalue matrix of $H_0$ 
but for simplification we use the former notation.
 As the $\tau$-dependence of $P$ 
in eq.(2) enters only through $\lambda_i$, a derivative of $P$ with respect 
to $\tau$ can be written as follows [10] 

\begin{eqnarray}
{\partial P({\mu_i},\tau})\over\partial\tau}= -\sum_{n=1}^N \int
{\prod_{i=1\not=n}^{N}\delta(\mu_i-\lambda_i)
{\partial \delta(\mu_n-\lambda_n)\over \partial \mu_n}
{\partial \lambda_n\over\partial\tau}
\rho (V){\rm d}V 
\end{eqnarray}

The further treatment of above equation depends on the symmetry-classes 
of the matrices $V$ and $H$, that is, whether they are real-symmetric or 
complex hermitian. 
	
	In the real-symmetric case, 
 ${\partial\lambda_n \over\partial\tau}$ can be expressed in terms of 
the matrix elements of the orthogonal matrix $O$ which diagonalizes $H$, 
$H=O^{T}.\Lambda.O$ with $\Lambda$ as the eigenvalue matrix
(Appendix A). Using this in eq.(3) leads to the following form with 
$g_{kl}=(1+\delta_{kl})$,

\begin{eqnarray}
{\partial P({\mu_i},\tau})\over\partial\tau}  
=\Omega^2\sum_{n}{\partial \over \partial\mu_n}\left(\mu_n P\right)
-2h^{-1}\Omega\sum_n  {\partial \over \partial\mu_n} \int
{\prod_{i}\delta(\mu_i-\lambda_i)
\sum_{k\leq l} O_{nk} O_{nl} {V_{lk}\over g_{kl}} \rho(V) {\rm d}V 
\end{eqnarray}

an application of the partial integration on the second term in the 
right of above equation then gives

\begin{eqnarray}
{\partial P({\mu_i},\tau)\over\partial\tau}  
 &=& \Omega^2\sum_{n}{\partial \over \partial\mu_n}\left(\mu_n P\right)
- h^{-1}\Omega\sum_n {\partial \over \partial\mu_n}
\int \sum_{k\leq l} {1\over \alpha_{kl}g_{kl}}
{\partial\over\partial V_{kl}}\left[\prod_{i}\delta(\mu_i-\lambda_i)
O_{nk} O_{nl}\right] \rho(V) {\rm d}V 
\end{eqnarray}

For simplification let us assume a same variance for all the  diagonal 
matrix elements such that $ g_{kl}\alpha_{kl} = \alpha$ for $k=l$.  
 Now by expressing 
$\sum_{k\le l} {1\over 
\alpha_{kl}g_{kl}}....={1\over \alpha}\sum_{k\le l}... -\sum_{k<l} 
\left({1\over\alpha}-{1\over g_{kl}\alpha_{kl}}\right)$ and with the help of 
 real-symmetric analog of relations (A3-A8) as well as the partial integration 
technique we can reduce  
eq.(5) as follows  

\begin{eqnarray}
{\partial P({\mu_i},\tau)\over\partial\tau} &=& 
\Omega^2\sum_{n}{\partial \over \partial\mu_n}\left(\mu_n P\right)
+{\Omega\over h\alpha}\sum_n {\partial \over \partial\mu_n}
\sum_m {\partial \over \partial\mu_m}
\int \prod_{i}\delta(\mu_i-\lambda_i) \sum_{k\leq l}
\left[{\partial\lambda_m\over\partial V_{kl}}{O_{nk} O_{nl}}
\right]\rho(V){\rm d}V \nonumber \\
&+& {1\over \alpha}\sum_{n,m,n\not=m} {\partial \over \partial\mu_n}
\int \prod_{i}\delta(\mu_i-\lambda_i) 
{\rho(V)\over {\lambda_m-\lambda_n}}{\rm d}V \nonumber\\ 
&-& h^{-1}\Omega \sum_{k<l}(g_{kl}^{-1}\alpha_{kl}^{-1}-\alpha^{-1})\sum_n 
{\partial \over \partial\mu_n}
\int  \left[ {\partial\over\partial V_{kl}}
\left(\prod_{i}\delta(\mu_i-\lambda_i){O_{nk}O_{nl}}\right)\right]\rho(V){\rm d}V 
%&+& {2\Omega^2\gamma'\over h^2\alpha'}\sum_n 
%{\partial \over \partial\mu_n}\int
%\prod_{i}\delta(\mu_i-\lambda_i)
%\left[\sum_{k \lt l}{\partial\lambda_n\over\partial V_{kl}}
% {V^3_{lk}\over g_{kl}}\right]\rho (V){\rm d}V  
%+ {2\Omega^2\over h^2}{\gamma\over \alpha}
%\sum_n {\partial \over \partial\mu_n}\int
%\prod_{i}\delta(\mu_i-\lambda_i)
%\left[\sum_k {\partial\lambda_n\over\partial V_{kk}}
% {V^3_{kk}\over g_{kk}}\right]\rho (V){\rm d}V  
\end{eqnarray}

By applying the orthogonality relation of matrix $O$, the eq.(6) can 
now be rewritten as follows

\begin{eqnarray}
{\partial P({\mu_i},\tau)\over\partial\tau} &=& 
\Omega^2\sum_{n}{\partial \over \partial\mu_n}\left(\mu_n P\right)
+{1\over \alpha}\sum_n {\partial \over \partial\mu_n}
\left[ {\partial \over \partial\mu_n} +
\sum_{m\not=n}{\beta \over {\mu_m-\mu_n}}\right] P- F 
\end{eqnarray}

where $\beta=1$ and

\begin{eqnarray}
F &=& h^{-2}\Omega^2 \sum_{k<l} \; (g_{kl}^{-1}\alpha_{kl}^{-1}-\alpha^{-1}) 
\int \prod_{i}\delta(\mu_i-\lambda_i) 
\left[\alpha_{kl} - 2 \alpha_{kl}^{2} V_{kl}^2\right]\rho(V) {\rm d}V 
\end{eqnarray}

Following the similar steps and the appropriate relations given in 
Appendix A, 
one obtains the same equation in complex hermitian case too (with $\beta=2$). 
The further analysis of the eq.(7) depends on the relative values of 
variances for the off-diagonal matrix elements. For a clear understanding, let 
us first consider a case where all the off-diagonals have the same variances,  
such that $g_{kl}\alpha_{kl}=\alpha'$ (for $k\not= l$). $F$ then can be expressed 
in terms of the derivative of $P$ with respect to 
$y\equiv {\alpha' \over \alpha}$, 

\begin{eqnarray}
F &=& 2h^{-2}\Omega^2 (1-y)y {\partial P\over\partial y}
\end{eqnarray}

Here we have used an idea quite important to our analysis 
that the variances of the matrix elements of $V$ can also vary and therefore 
eigenvalues evolve not only as a function of the perturbation strength but 
the variances too. 
Note here  that $F$ can also be expressed as an extra drift term
($F= \sum_n {\partial R(\mu)\over \partial \mu_n}$),
with $R= \int \left[\sum_{k\le l}(g_{kl}^{-1}\alpha_{kl}^{-1}-\alpha^{-1})
{\partial\over\partial V_{kl}}
\left(\prod_{i}\delta(\mu_i-\lambda_i){O_{nk}O_{nl}}\right)\right]
\rho (V){\rm d}V$   and it can be of 
interest to study the effect of this extra drift on the dynamics. For example, 
for $R$ nonlinear in $P$, the motion may loose its random behaviour, showing 
some sort of periodicity. But in this paper, we restrict ourselves just to  
explore the universality of the dynamics, that is, to verify that $P$ satisfies a 
similar F-P equation as the one proposed by Dyson.

.

The eq.(7) contains the derivatives of $P$ with respect to 
two parameters, namely, $\tau$ and $y$. In order to prove that the 
statistical quantities in the GG case evolve in a similar way
as in the SG case, 
 one should be able to reduce the two parameteric dependence  
of eq.(7) in a single parameter.
We assume that it is possible along a curve
parameterized by $\phi$, $\tau=\tau(\phi), y=y(\phi)$
 such that

\begin{eqnarray}
{\partial\over \partial\phi} = 
{\partial \tau\over \partial\phi}{\partial\over \partial\tau}
+ {\partial y\over \partial\phi}{\partial\over \partial y}
\end{eqnarray}

A comparison of eq.(10) with eq.(7) then leads us to the conditions  
for existence of such a curve 
 namely, 
${\partial \tau\over \partial\phi} = 1$ and  
${\partial y\over \partial\phi} = {2\Omega^2 y(1-y)\over h^2}$.
By solving these equations, one obtains the following curve in the $(y,\tau)$-space,

\begin{eqnarray}
y &=& {{{\rm e}^{2\Omega^2 \tau}-1}\over {{\rm e}^{2\Omega^2 \tau}+y_0}}
\qquad {\rm and} \quad\tau = \phi + \tau_0 
\end{eqnarray}

where $\tau_0$ and $y_0$  are arbitrary constants. 
Note the
conditions mentioned above do not impose any constraints on these
constants except that the curve should remain on
the positive side of the upper-half of $y-\tau$ plane. 
The extraction of $\tau_0$ from eq.(11) 
leads to the following form of $\phi$

\begin{eqnarray}
\phi=\tau -{1\over 2\Omega^2}{\rm log}\left[1+d {|y-1|\over y}
({\rm e}^{2\Omega^2\tau}-1)\right] 
\end{eqnarray}
 
where $d={y_0\over y_0-1}$ and can be chosen as unity (as the corresponding 
value of $\phi$ still satisfies the required conditions). 
Thus, for $y={\alpha'\over\alpha}$, eq.(11) represents a set of 
curves along which eq.(7) adopts the form of a F-P equation

\begin{eqnarray}
{\partial P({\mu_i},\phi)\over\partial\phi} &=& 
\Omega^2\sum_{n}{\partial \over \partial\mu_n}\left(\mu_n P\right)
+{1\over \alpha}\sum_n {\partial \over \partial\mu_n}
\left[ {\partial \over \partial\mu_n} +
\sum_{m\not=n}{\beta \over {\mu_m-\mu_n}}\right] P
\end{eqnarray}
here the steady state is achieved for $\phi \rightarrow
\infty$ which corresponds to $\tau \rightarrow \infty$ and
$\alpha \rightarrow \alpha'$; the steady state solution is given by  
$\prod_{i<j} |\mu_i-\mu_j|^{\beta}
{\rm e}^{-\alpha\Omega^2\sum_k \mu_k^2}$. Note that only $\tau \rightarrow 
\infty$ 
(with ${\partial P\over \partial \tau}=0$) no longer represents the 
steady state as in SG or WD case but represents a 
transition state with $\phi (y) =-{1\over 2\Omega^2}{\rm log}( d {|y-1|\over y})$.

The steady state limit of $P$ at $(y,\tau)=(1,\infty)$ seems to indicate that
if the initial perturbation is such that the width
($\propto {1\over \alpha}$) of the eigenfunctions of $H$ is
smaller than the overlapping
between them ($\propto {1\over \alpha'}$), they have a
tendency to overcome this difference and extend over whole of the 
available Hilbert space. This also 
seems to suggest that a complicated interaction giving rise to larger transition 
probabiliies between energy levels as compared to the probability of 
staying in the same level can only be transient in nature. Given freedom 
to change, it rearranges itself in such a way so as to equalize these 
probabilities.

This can be seen directly from 
eqs.(7,8) because in $N\rightarrow\infty, \Omega \rightarrow 0$ 
and $H\rightarrow V$ thus removing any $\Omega$-dependence of $\lambda_i$'s  
and making $F \rightarrow 0$. A same result, for SG type perturbation in large 
$N$-limit,
has already been obtained [9]. The analogous behaviour in both GG 
and SG case can be attributed to the  
negligible effect of the distribution of $N$-diagonal matrix 
elements on the dynamics in presence of $N(N-1)/2$ off-diagonal 
ones. This indicates that, in the thermodynamic limit, difference of 
the variances  does not leave its trace on the evolution of eigenvlaues.

        Again, in $N\rightarrow \infty$ limit, all the three  
distributions, namely, SG, GG and WD evolve in the same way for
arbitrary initial conditions. Thus all moments of the eigenvalues  
in SG and GG cases at any value of the transition parameter will be   
equal to the corresponding moments of particle positions of a
WD gas undergoing Brownian motion due to the presence of
thermal noise. However, as discussed in [9], this does not imply a
Brownian motion of eigenvalues for the first two case. This is due to 
the difference of sources randomizing the motion of eigenvalues. For
SG and GG cases, the source is the matrix $V$, acting like a quenched disorder 
while, for WD gas, the thermal noise  gives rise to an annealed randomness.
Although these different origins of randomness do not affect the
static (for one parameter value) as well as the $2^{nd}$-order
parametric correlations but, as shown in ref.[9],
the multiple parametric correlations higher than the $2^{nd}$-order are
different for the perturbed systems and WD gas.  However, the 
 origin of randomness being similar for the SG and GG cases, all the
 conclusions obtained for the correlations for the former [9]
 are also valid for the latter in the thermodynamic limit.

It is interesting to note that  eq.(4) can also be written 
in the following form
(by first using the real-symmetric analog of eq.(A3) in eq.(4) and then taking  
the derivative with respect to $\mu_n$ inside the integral) 

\begin{eqnarray}
{\partial P({\mu_i},\tau)\over\partial\tau} &=& 
 \Omega^2\sum_{n}{\partial \over \partial\mu_n}\left(\mu_n P\right)
+ h^{-2}\Omega^2 \sum_{k\leq l}  \int   
{\partial  {\prod_{i}\delta(\mu_i-\lambda_i)}
\over \partial V_{kl}} V_{kl} \rho(V) {\rm d}V 
\end{eqnarray}

Now an application of the partial integration to the second term on the 
right of this equation gives

\begin{eqnarray}
\int {\partial  {\prod_{i}\delta(\mu_i-\lambda_i)}
\over \partial V_{kl}} V_{kl} \rho(V) {\rm d}V 
= \int {\prod_{i}\delta(\mu_i-\lambda_i)} 
\left[ 1 + 2 \alpha_{kl} V_{kl}^2 \right] 
\rho(V) {\rm d}V 
\end{eqnarray}

By expressing this in terms of derivatives of $P$ with respect to 
$\alpha, \alpha'$ as before, we obtain a drift equation 
(without a diffusion term as well as the repulsive pairwise potential)

\begin{eqnarray}
{\partial P({\mu_i},\tau)\over \partial\phi_1} 
=\Omega^2\sum_{n}{\partial \over \partial\mu_n} (\mu_n P)
\end{eqnarray}

where $\phi_1$ describes a curve in a 3-dimensional ($y,\alpha,\alpha'$)  
space such that 
${\partial P\over \partial\phi_1} = {\partial P\over \partial\tau} +
{2\Omega^2 \over h^2} \left[\alpha' {\partial P\over \partial\alpha'} + 
\alpha {\partial P\over \partial\alpha}\right]$. The above possibility 
of reduction of the F-P equation in a pure drift equation just by going to a 
higher parametric space implies the total blindness of eigenvalues to any 
mutually repulsive potential or locally fluctuating force in this space. Note that  
it is already well-known that a repulsion between the 
eigenvalues vanishes if the number of parameters of the hamiltonian 
undergoing variation is more than 
 one in real-symmetric case (two for complex-hermitian case) [11]. However 
the same situation seems to exist even when the parameters undergoing variation 
 are the ones governing the distribution of matrix elements of hamiltonians. 

	So far we have considered the case where all the off-diagonal 
	matrix elements $V_{ij}$ have same variance although different 
from that of diagonal. But in many physical situations e.g. in disordered 
systems, there occurs localization of eigenfunctions which may give rise 
to  off-diagonal elements with different variances. One such situation most 
often encountered is  where $V_{ij}$'s decay exponentially with 
respect to the distance from diagonal. It can be modeled by 
taking different variances for different off-diagonals, that is, 
$g_{kl}\alpha_{kl} = \alpha_r, \quad   \qquad  (r=|k-l|)$. 
 $F$ in such a case is given by, with $y_r \equiv {\alpha_r \over \alpha}$ 

\begin{eqnarray}
F = 2 h^{-2}\Omega^2\sum_{r=1}^{N-1}
(1-y_{r})y_r {\partial P\over\partial y_{r}} 
\end{eqnarray}

We assume that it is possible along a curve
parameterized by $\phi$, $\tau=\tau(\phi), Y=Y(y_1,..,y_N)$
 such that

\begin{eqnarray}
{\partial\over \partial\phi} = 
{\partial \tau\over \partial\phi}{\partial\over \partial\tau}
+ {\partial Y\over \partial\phi}{\partial\over \partial Y}
\end{eqnarray}

with ${\partial\over \partial Y} = \sum_{r}
{\partial y_r\over \partial Y}{\partial\over \partial y_r}$.
Reasoning again as before one finds that the universality of evolution 
of the probability density still survives but only by going to a 
higher ($N$-dimensional) 
parametric space $(\tau,y_1,..,y_N)$, along a set of curves 
parametrized by $\phi$ and given by the conditions 
${\partial \tau\over\partial \phi} =1 \quad 
{\partial Y\over\partial \phi}= 2 h^{-2}\Omega^2 $ with  
${\partial y_r\over\partial Y}=  (1-y_r)y_r$. Again    
$\phi$ can be obtained by solving these equations,

\begin{eqnarray}
\phi=\tau -{1\over 2\Omega^2}{\rm log}\left[1+d \prod_{r=1}^N 
  \left({|y_r-1|\over y_r}\right)
({\rm e}^{2\Omega^2\tau}-1)\right] 
\end{eqnarray}
Note here 
that $\prod_r$ contains the conribution only from those $y_r$'s for which 
$y_r-1\not=0$.

	One can also consider a physical 
situation when all the off-diagonal matrix elements have different 
probability laws $g_{kl}\alpha_{kl} = \alpha, ({\rm for}\; k=l)$ and 
$\alpha_{kl} \not= \alpha_{ij}$  if $({kl} \not= {ij})$. In this case,  
$F$ now turns out to be 
as follows with $y_{kl}\equiv {\alpha_{kl}\over \alpha}$,

\begin{eqnarray}
F = 2 h^{-2}\Omega^2\sum_{k<l}(1-y_{kl})y_{kl}
 {\partial P\over\partial y_{kl}} 
\end{eqnarray}

Now one has to go to ${N(N-1)\over 2}+1$-dimensional parametric 
space to recover the F-P equation similar to Wigner-Dyson gas. 
The curves in this case are again parameterized by $\phi$
 which is still related to $\tau$ and $Y$ by eq.(18) but now   
 $Y\equiv Y(\{y_{kl}\}) $ is such that 
${\partial\over \partial Y} = \sum_{k<l}
{\partial y_{kl}\over \partial Y}{\partial\over \partial y_{kl}}$
with
${\partial y_{kl}\over \partial Y}=(1- y_{kl})y_{kl}$; the two other 
derivatives of $\tau$ and $Y$ with respect to $\phi$ remain the same 
as in the above case. 
Proceeding exactly as before, $\phi$ in this case can be shown to be 
following, 

\begin{eqnarray}
\phi=\tau -{1\over 2\Omega^2}{\rm log}\left[1+d \prod_{k<l}^N 
  \left({|y_{kl}-1|\over y_{kl}}\right)({\rm e}^{2\Omega^2 \tau} -1)\right]
\end{eqnarray}
Again, as before, the $\prod_{k<l}$ contains 
contributions from all $y_{kl}\not=1$.

For SG case, $\phi$ appearing 
in the correlators of the type $<d(E,\phi),d(E,0)>$  involves variation of 
just one parameter, for CG case, two parameters or more. However, in both the cases,  
the expression of second order correlators is same due to same form of the F-P equation. 
 Our study therefore suggests that the 
second order correlators involving variation of more than one parameter can always be 
expressed as those involving just one effective parameter. 
 For example the correlators belonging to RM models of both 
localized as well as delocalized case can be expressed in the same form by 
using parameter $\phi$ although the definition of $\phi$ is different in 
two cases. 

	Our study also suggests that nature of the dynamics of the eigenvalues  
is very sensitive to the number of parameters undergoing a change. For 
finite $N$, the motion may appear as pure drift suggesting no interaction 
between eigenvalues and a total absence of local fluctuating and frictional  
forces if variances of all the matrix elements  are allowed to 
vary independently. 
It seems Brownian as a function of just the relative variations of the 
variance. Under variation of only perturbation strength parameter, 
 the presence of an exta drift 
like term 
makes the motion non-Brownian. But the contribution of this term being zero in 
$N\rightarrow\infty$ limit, it recovers its Brownian nature.  
It can be of 
interest to study the effect of this extra drift on the dynamics.

  	Although the study presented here deals with the energy 
level-dynamics of random matrix models of hermitian operators e.g. 
hamiltonian of complex systems, the results
 are also valid for the
level-dynamics of unitary operators $U$ e.g
time-evolution operator being perturbed by a
random matrix $V$ taken from the GG ensembles,  
 $(U(\tau)=U_0 {\rm exp}[i \sqrt{\tau} V] \approx
U_0[1+i\sqrt{\tau} V]$ for small $\tau$ values). 
The required F-P equation in this case turns out to 
be the following

\begin{eqnarray}
{\partial P({\mu_i},\phi)\over\partial\phi} &=& 
{1\over \alpha}\sum_n {\partial \over \partial\mu_n}
\left[ {\partial \over \partial\mu_n} +{\beta\over 2}
\sum_{m\not=n}{\rm cot}\left({\mu_m-\mu_n \over 2}\right)\right] P
\end{eqnarray}
with $\phi$ defined as before. To
prove the equivalence for $U$ a symmetric unitary
matrix $(\beta=1)$, one has to follow the similar steps as in
the real-symmetric case of hermitian matrices as
both are invariant under orthogonal
transformation.  Similarly for a general unitary
$U$ $(\beta=2)$, the steps are similar to the complex
hermitian case. As obious from eq.(22), this is similar to the 
F-P equation obtained if $V$ belonged to the SG ensembles [12].
.
%.
%.

\section {Calculation of $P(\mu,\tau)$}

In the preceeding section, we obtained a F-P equation governing the evolution of 
probability $P(\mu,\tau)$ for an arbitrarily given $H(\phi_0)$ (the conditional 
probability). Here $\phi_0$ corresponds to the case $\tau=0$ and and $H(\phi_0)$ can be 
obtained by using the relation $H(\phi)=H(\phi_0){\rm e}^{-\Omega^2 (\phi-\phi_0)}+ 
\hat V \sqrt{1-{\rm e}^{-2\Omega^2(\phi-\phi_0)}}$, substituting $\phi$ in terms of 
$\tau$  
and $y$ there and then comparing it with the expression for the hamiltonian $H(\tau)$ 
given by eq.(1). Here $\hat V$ is a matrix with an invariant probability distribution 
($\rho(\hat V) \propto {\rm e}^{-\alpha {\rm Tr}{\hat V}^2}$) and $H(\phi)$ can be 
written in the above form as it gives the correct evolution equation for the 
probability distribution. 
 
The formal similarity of the F-P equation with that of WD 
gas case as well as SG case makes it easier to obtain the $P(\mu,\tau)$ at least 
for $\beta=2$ case as the solution for the latter case is already known. The 
$P(\mu,\phi)$ (and therefore $P(\mu,\tau)$ can also be calculated directly 
from hamiltonian 
$H(\phi)$ by using sum of the matrices technique (that is, by evaluating a two-
matrix integral) as for the SG case [4] because now both the component matrices 
have invariant distribution unlike eq.(1). 
For completeness purpose, We give here few of the steps, used in solving SG case, 
for our case; for details refer to [12]. 

        As can readily be checked, both eq.(13) and (22) can also be written as follows    
(with $\alpha=1$ for simplification)
\begin{eqnarray}
{\partial P({\mu_i},\phi)\over\partial\phi} &=& 
\sum_{n}{\partial \over \partial\mu_n}|Q_N|^{\beta} 
{\partial \over \partial\mu_n}{P\over |Q_N|^{\beta}} 
\end{eqnarray}

where $|Q_N|^{\beta}=|\Delta (\mu)|^{\beta}
{\rm e}^{-\omega^2 \sum_k \mu_k^2}$
with $\Delta (\mu)=\prod_{i<j}(\mu_i-\mu_j)$
for the hermitian case and $=\prod_{j<k}{\rm sin}\left(\mu_j-\mu_k \over 2\right)$ 
for the unitary case. 
The transformation 
$\Psi=P/|Q_N|^{\beta/2}$ allows us to cast eq.(23) in the suggestive form

\begin{eqnarray}
 {\partial \Psi \over\partial\phi} = -\hat H \Psi 
\end{eqnarray}
where, for the hermitian case, the 'Hamiltonian' $\hat H$ turns out to be the  
Calogero-Moser Hamiltonian  
\begin{eqnarray}
\hat H &=& 
\sum_i{\partial^2 \over \partial\mu_i^2}
-{1\over 2}\sum_{i<j}{\beta (\beta-2)\over (\mu_i-\mu_j)^2}+
{\Omega^4 \over 4}\sum_i 
\mu_i^2 
\end{eqnarray}
%\narrowtext 
Similarly, for the unitary case, $\hat H$ is the Calogero-Sutherland hamitonian
 
\begin{eqnarray}
\hat H &=& 
-\sum_i{\partial^2 \over \partial\mu_i^2}
+{\beta(\beta-2)\over 16}\sum_{i\not=j}{\rm cosec}^2 \left(\mu_i-\mu_j\right)
-{\beta^2\over 48} N(N^2-1)
\end{eqnarray}

With the parabolic-confining potential (or periodic boundary conditions) 
and under the requirement (to take account of
the singularity in $H$) that the solutions vanish as $|\mu_i-\mu_j|^{\beta/2}$ 
when $\mu_i$ and $\mu_j$ are close to each other, $\hat H$, both in eq.(25) and (26), 
 has well-defined (completely 
symmetric) eigenstates $\zeta_k$ and eigenvalues $\lambda_k$. 
This allows us to express 
the "state" 
$\Psi$ and therefore $P(\mu,\phi| H_0)$ as a sum over eigenvalues and eigenfunctions 
of $\hat H$

\begin{eqnarray}
P(\mu,\phi|H(\phi_0))=|{Q_N(\mu) \over Q_N(\mu_0)}|^{\beta/2}
 \sum_{k>0} {\rm exp}[-\lambda_k (\phi-\phi_0)]\zeta_k(\mu)\zeta_k^*(\mu_0)
\end{eqnarray}

where $\mu_0\equiv (\mu_{01},\mu_{02},...,\mu_{0N})$ are the eigenvalues of $H(\phi_0)$. 
%. 
The joint probability distribution $P(\mu,\phi)$ can then be obtained by integrating 
over all initial conditions

\begin{eqnarray}
P(\mu,\phi)= \int P(\mu,\phi|\mu_0,\phi_0)P(\mu_0,\phi_0) {\rm d}\mu_0
\end{eqnarray}

which further leads to correlations using standard techniques [12,13]. 

	The eqs.(27-28) represent the formal solutions of eq.(13) (or eq.(22)). 
   To proceed further, one needs to know the eigenvalues and eigenfunctions of $\hat H$ 
so as to express $P$ in a compact form but these are explicitly known only for 
  $\beta=2$ case [8]. 
 This is because for $\beta=2$ the interaction 
term in eqs.(25-26) drops out and then $P$ can be obtained explicitly.
For CM model (eq.(25)), it is given as follows [4],
\begin{eqnarray}
P(\mu,\phi|H(\phi_0);\beta=2) \propto 
 |{\Delta(\mu) \over \Delta(\mu_0)}|
{\rm det}[f_m(\mu_i-\mu_{0j};\phi-\phi_0)]_{i,j=1...N}
\end{eqnarray}
with $f_m(x-y;t)={\rm exp}\left[-{x{\rm e}^{\Omega^2 t}-y \over 
{\rm e}^{2\Omega^2 t}-1}\right]$,
and, for CS model [12], 
\begin{eqnarray}
P(\mu,\phi|H(\phi_0);\beta=2) = 
{1\over N{\rm !}} |{\Delta(\mu) \over \Delta(\mu_0)}|
{\rm det}[f_s(\mu_i-\mu_{0j};\phi-\phi_0)]_{i,j=1...N}
\end{eqnarray}
where $f_s(x)={1\over 2\pi}\sum_{k=-\infty}^{\infty}{\rm exp}[-k^2 \phi +i k x]$.

As we already 
know $P(\mu,\phi)$ and various correlations for WD case (for which $\phi=\tau$) starting 
from various initial conditions [12,13], one can obtain these measures for GG case 
just by substituting $\phi$ by its appropriate relationship with $\tau$ and variances of 
the perturbation-matrix elements. For example, for the case $H=V\Omega^{-1}$ (which corresponds 
to $H(\tau\rightarrow\infty)$ in eq.(1)) with different variances for different diagonals of $V$ 
(eq.(17)), $\phi=-(2\Omega^2)^{-1}{\rm log}[\prod_{r=1}^N {|y_r-1|\over y_r}]$ (with $d$ 
chosen to be unity for the same reason as in eq.(12)). Also 
note that, in this case, ${\partial P\over \partial \tau}=0$. The initial condition 
$\phi=0$ now corresponds to all $y_r \rightarrow\infty$, thus implying the existence 
of only the diagonal elements of $V$ and a Poisson distribution for $P(\mu_0,\phi=0)$ 
 $(\propto {\rm e}^{-\Omega^2 \alpha\sum_i V_{ii}^2}$ with $V_{ii}=\mu_{0i}$). As 
$\phi\rightarrow\infty$ when $y_r \rightarrow 1$ for all $r$, the  equilibrium  
distribution is therefore given by the standard Gaussian ensembles. This case thus 
corresponds to the Poisson $\rightarrow$ GE transition in the standard Gaussian 
ensembles [12,13] with $\phi$  
 now as a transition parameter (replacing $\tau$, the perturbation strength) with 
intermediate ensembles representing the cases 
for any choice of $y_r$'s. The two-point correlations for this transition 
 for $V$ belonging to the SG ensemble 
(complex-hermitian) have already been obtained [13]. It should be noted that, as for  
 the SG ensembles,
$\phi$ must be rescaled to see the smooth transition [14].  
.

\section{Polynomial Case}

	In section II, we showed the equivalence of the dynamics 
of the Wigner-Dyson gas with the evolution of the eigenvalues of a 
hamiltonian under a perturbation $V$ taken from a generalized Gaussian 
ensembles. However the claim about universality of dynamics and density 
correlators can only be made if 
the equivalence is proved, at least in thermodynamic limit, for a $V$ with 
distribution of more general nature. In this section, We attempt to verify 
the claim by taking $\rho (V)$ as the exponential of the 
polynomial form of $V$,  
 $\rho (V)=C{\rm exp}({-\sum_{k\leq
l}  Q(V_{kl})
})$ with $Q(x)=\sum_{r=1}^M \gamma_{kl}(r)x^{2r}$ (a polynomial 
of $x$ with degree $2M$),  
$C$ as the normalization constant
(referred as polynomial case I later on) and 
variances for diagonal and off-diagonal matrix
elements chosen to be arbitrary. Note the universality of the correlators 
for the case $H=V$, with $V$ distributed as in this case, has already 
been studied in Ref.[14] (which corresponds to the steady state limit 
of the study 
given here) 

To obtain a F-P equation, we now need following equality
which can be proved by using integration by
parts,

\begin{eqnarray}
\int f[V]\; V_{kl} \;\rho(V)\; {\rm d}V_{kl} 
&=& {1\over 2\gamma_{kl}(1)}\int{\partial f[V] \over
\partial V_{kl}}
\rho(V) {\rm d}V - \sum_{r=2}^k r{\gamma_{kl}(r)\over \gamma_{kl}(1)}
\int V^{(2r-1)}_{kl} f[V]\rho(V) {\rm d}V 
\end{eqnarray}

By using the real symmetric analog of eq.(A1) in eq.(3), followed by above equality, we get

\begin{eqnarray}
{\partial P\over\partial\tau} &=&
\Omega^2\sum_{n}{\partial \over \partial\mu_n}\left(\mu_n P\right)
-h^{-1}\Omega\sum_n {\partial \over
\partial\mu_n}
\int \sum_{k\leq l} {1\over \gamma_{kl}(1) g_{kl}}
{\partial\over\partial V_{kl}}
\left[\prod_{i}\delta(\mu_i-\lambda_i)
O_{nk} O_{nl}\right] \rho(V) {\rm d}V \nonumber\\
&+& 2h^{-1}\Omega\sum_{r=2}^M r\sum_n {\partial \over
\partial\mu_n}\int\prod_{i}\delta(\mu_i-\lambda_i)
\left[\sum_{k\leq l} {\gamma_{kl}(r)\over \gamma_{kl}(1)} 
{V^{2r-1}_{kl}\over g_{kl}}
O_{nk} O_{nl} \right] \rho(V) {\rm d}V
\end{eqnarray}

Now for simplification let us consider a case
where 
$g_{kl}\gamma_{kl}(r)=\gamma(r)$ if $(k=l)$ and $
g_{kl}\gamma_{kl}(r) =\gamma'(r)$ if $(k\not=l)$.  

Using the tools given in appendix A and by an extensive
use of the integration by parts while dealing
with nasty integrals, 
the first term appearing in above equation can now be 
reduced in the same form as in GG case. This results 
in the following

\begin{eqnarray}
{\partial P(\{\mu_i\},\tau)\over\partial\tau} &=&
\Omega^2\sum_{n}{\partial \over \partial\mu_n}\left(\mu_n P\right)
+{1\over \gamma'(1)}\sum_n {\partial \over
\partial\mu_n}
\left[ {\partial \over \partial\mu_n} +
\sum_{m\not=n}{\beta\over {\lambda_m-\lambda_n}}\right] P-Z
\end{eqnarray}
where $\beta=1$ and
\begin{eqnarray}
Z &=& {\Omega^2\over
 h^2}({1\over\gamma(1)}-{1\over\gamma'(1)})F
-{\Omega^2\over h^2}\sum_{r=2}^k r \left[{\gamma'(r)\over\gamma'(1)}G_{r1}
+{\gamma(r)\over\gamma(1)}G_{r2}\right]
\end{eqnarray}

with $F,\; G_{r1},\;G_{r2}$ given as follows
\begin{eqnarray}
F &=& {1\over 2}\sum_n {\partial \over \partial\mu_n}
\int \left[\sum_{k} {\partial\over\partial V_{kk}}
\left(\prod_{i}\delta(\mu_i-\lambda_i)
{\partial\lambda_n\over\partial V_{kk}}g_{kk}
\right)\right]\rho (V){\rm d}V \\
G_{r1} &=& \sum_n {\partial \over \partial\mu_n}\int
\prod_{i}\delta(\mu_i-\lambda_i)
\left[\sum_{k<l}{\partial\lambda_n\over\partial V_{kl}}
 V^{2r-1}_{kl}\right]\rho (V){\rm d}V \\ 
G_{r2} &=& \sum_n {\partial \over \partial\mu_n}\int
\prod_{i}\delta(\mu_i-\lambda_i)
\left[\sum_k {\partial\lambda_n\over\partial V_{kk}}
 V^{2r-1}_{kk}\right]\rho (V){\rm d}V
\end{eqnarray}

We apply partial integration to F and rewrite it
as follows,

\begin{eqnarray}
F &=&\gamma(1)\sum_k \; \int \prod_{i}\delta(\mu_i-\lambda_i) 
\left[1 -  \gamma(1)  V_{kk}^2 - \sum_{r=2}^M r \gamma(r) V^{2r}_{kk}
\right]\rho(V) {\rm d}V  + \sum_{r=2}^k r\gamma(r) G_{r2} 
\end{eqnarray}

similarly $G_{r1}$ and $G_{r2}$ can be rewritten as follws,

\begin{eqnarray}
G_{r1} &=& \sum_{k<l} \int \prod_i\delta(\mu_i - \lambda_i)\;
\left[(2r-1) V_{kl}^{2r-2} - 2 \gamma'(1) V_{kl}^{2r} - 
2\sum_{s=2}^M s\gamma'(s) V_{kl}^{2r+2s-2}
\right] \;\rho (V){\rm d}V  
\end{eqnarray}

\begin{eqnarray}
G_{r2} &=& \sum_{k} \int\prod_i\delta(\mu_i - \lambda_i)\;
\left[(2r-1) V_{kk}^{2r-2} - \gamma(1) V_{kk}^{2r} - 
\sum_{s=2}^M s \gamma(s) V_{kk}^{2r+2s-2}
\right] \;\rho (V){\rm d}V  
\end{eqnarray}

For the complex Hermitian case also, one obtains
an equation similar to eq.(33) with $\beta=2$ and $Z$ 
given by eq.(34). 
Note that all the terms appering in the 
expressions of $F, G_{r1}, G_{r2}$ are of the type 
$\int \prod_i (\mu_i-\lambda_i) V^{j} \rho(V) {\rm d}V $, 
and, as done in section II for GG case, they can easily 
be rewritten in terms of 
the derivatives with respect to $\gamma (r)$'s if $j \le 2M$. 
It is difficult to do so for terms with $j>2M$, they probably  
could be expressed as higher order derivatives with respect to 
more than one variance parameters (this is the case at least for 
$N=2$). 
However as physically significant RM models of complex systems 
generally correspond to limit $N\rightarrow \infty$, it would be 
sufficient, for our purpose, to study the behaviour of the terms 
in this limit. 
As obvious from eqs.(38-40), the $N$ or $\Omega$-dependence of $F, G_{r1}$ and $G_{r2}$ 
is  due to  presence of
the $\lambda_i$'s which can be clearly   
seen by using the equality $\delta(\lambda^{-1}[\mu]-V)=\delta (\mu-\lambda[V]){\rm det}
|{\partial \lambda \over \partial V}|$ and rewriting these integral in terms of the 
function $\lambda^{-1}(\mu)$.
Let us consider the case when both $\gamma (r)$ and 
$\gamma (r')$, $r=1 \rightarrow M$, are independent of $N$ and all 
the matrix elements of $V$ are distributed with zero mean; the latter   
choice implies the N-independence of $V_{kl}$ also.
Thus the $N$-dependence of $F, G_{r1}$ and $G_{r2}$ in this case  
 is of the same order as appears in $P$   
and therefore it follows from
eq.(34) that the contribution from $Z$ to  eq.(33) is negligible (as compared to 
the diffusion term and the drift term due to mutual repulsion) in that limit. 
 Similarly for GG case, $Z$ vanishes for $N\rightarrow \infty$; $Z$
here can be obtained by substituting
$\gamma(r)=0,\;\gamma'(r)=0$, for $r>1$, in eq.(13).  Further for
the case where distribution, although basis
independent, contains a non-Gaussian term, that
is, for $\rho(V) \propto exp[-\alpha\; Tr (V^2)
-\sum_{r=2}^M\gamma(r)\; Tr(V^{2r})]$ (polynomial case $II$), the $Z$ can be
shown to be following,

\begin{eqnarray}
Z &=& {2\gamma\Omega^2 \over h^2}
\sum_{r=2}^k r \gamma(r) \nonumber \\
& &\int\prod_i\delta(\mu_i - \lambda_i)\;
\left[\sum_{k,l} g_{k,l} {\partial (V^{2r-1})_{kl}\over\partial V_{kl}}
 - 4\alpha Tr(V^{2r}) - 4 \sum_{s=2}^k Tr(V^{2r+2s-2})
\right] \;\rho (V){\rm d}V  
\rightarrow 0 \quad {\rm for} \quad
N\rightarrow \infty
\end{eqnarray}

In thermodynamic limit ($N \rightarrow \infty$),
the eq.(33) is same as the F-P equation
governing the Brownian motion of particles in
Wigner-Dyon gas (also the Sutherland model) with
particle positions and time in the latter
replaced by $\mu_i$ and $\tau$ in the former. This 
remains valid also for the case when $Q(x)$ is chosen 
to be an arbitrary function expandable in a Taylor series. 
Thus we find that, under a perturbation taken from a 
ensemble with a sufficiently general distribution, the distribution of eigenvalues
of the quantum system 
evolves in the same way as the distribution of
particle position in the Wigner-Dyson gas for
arbitrary initial conditions. 
 This also implies
that all the moments of the eigenvalues
calculated for a parameter value $\tau$ will be
equal to the corresponding moments of particle
positions of the latter. But the motion of the
eigenvalues is not Brownian because the
randomness in their motion comes from the matrix
$V$ which acts like quenched disorder [9]. 
 
This also implies the equivalence of second order
parametric correlators in the two cases because
they can be expressed as a sum over various
moments of eigenvalues, weighted suitably and
then averaged over equilibrium initial
conditions. However, the F-P equation being Markovian
in nature, its equivalence for the two cases can not
lead us to a similar conclusion for the 
the higher order correlators which involve moments
at more than one parameter value and therefore multiple
parameter averaging. As mentioned in [9], 
the $n$-point correlations for the Wigner-Dyson gas 
can be expressed in terms of the two point functions, 
yielding an $n$-matrix integral while for RM models 
of quantum chaos, the $n$-point function remains a 
two matrix integral.

      It shuold be noted here that 
we have considered the case only for $N$-independent 
coefficients in the polynomial; a more careful analysis is 
required when these have different $N$-dependence. For example, 
if one or more coefficients are $N$-dependent, increasing with $N$ 
increasing, the contribution to eq.(33) from terms in eqs.(38-40) 
is no longer negligible and therefore the evolution of $P$ is no 
longer the same as in Gaussian case.

\section{Conclusion}

	In this paper, we have analytically studied the response of energy 
levels of complex quantum systems to external perturbations modeled by 
generalized random matrix ensembles. Our results indicate the universality 
of two-point parametric density correlators as well as the static 
correlators of all orders thus agreeing with the 
numerically observed results for complex systems. One interesting feature 
revealed by our study is that for both the localized and the delocalized quantum 
dynamics of these systems, the second order parametric correlations can be 
shown to have the same form except that the definition of the effective parameter  
is different in the two cases. The method adopted here 
handicaps us from making any statement about universality of the higher 
order parametric correlations.

	Further, for the reasons given in the sections (I,IV), the correspondence shown 
between Wigner-Dyson gas and the quantum 
hamiltonian $H=H_0 +x V$ with $V$ having a sufficiently general nature 
 also implies the equivalence between the space-time 
correlators 
 of the Calogero-Sutherland system and the second order parametric correlations 
of $H$.
 This further strengthens the idea contained in [1] and supported by studies 
in [9], namely, the  
$1$D Sutherland model provides a model  hamiltonian for the dynamics of 
eigenvalues of quantum chaotic systems [10].
	The equivalence shown between the F-P equation governing the evolution, of 
eigenvalues in GG and SG cases also turns out to be quite 
helpful in studying the correlations for various difficult but physically significant 
situations (depending 
on the variances of perturbation matrix) in the former case, by using the technique 
given in section III. Note that our result is quite general as we have shown the 
equivalence for arbitrary choice of variances. 
	As already mentioned in section II,  
the results obtained  in sections II, III 
and IV are also valid for the
level-dynamics of unitary operators $U$. 
 Note the analogy between the
statistical properties of non-equilibrium
Circular ensembles which model the eigenvalue
spectra of unitary operators and non-equilibrium
standard Gaussian ensembles has already been
proved [12].

	Our study still leaves many important questions unanswered. For 
example, universality or its absence among higher order correlations 
of complex systems and their similarities or differences with 
Wigner-Dyson gas is not fully understood. Further the results here have been 
obtained for explicit averaging over a generalized random perturbing 
potential. It would be interesting to know whether similar conclusions can 
also be obtained  for the complex systems where ensemble averaging is not valid 
(e.g. Billiards) and eigenvalue statistics should be computed as an average over 
the $N$ eigenvalues. The intution suggests a positive answer as the  
sufficiently complex systems  are quite likely to be self-averaging. This implies  
that a single choice of matrices from a particular distribution 
are representative in the large $N$-limit of the entire distribution and 
therefore an average over eigenvalues should give the same result as an 
ensemble average.

\acknowledgments
I am grateful to Prof. B.S.Shastry for suggesting me this study and for 
various useful ideas. I would also like to express my gratitude to 
 Prof. B.L.Alshulter and 
Prof. V.Kravtsov, the discussions with whom  have also helped in 
 improving this work. 
I am also obliged to JNCASR, India  for 
economic support during this study.   

\appendix
\section{}

For $H$ a complex hermitian matrix, its
eigenvalue equation is given by $\Lambda=
U.H.U^{+}$ with $\Lambda$ as eigenvalue matrix
and $U$ the eigenvector matrix which is unitary.
Now for a complex hermitian perturbation $V_{ij}
= V_{ij;1} + i V_{ij;2} = V^{*}_{ji}$ (with
$V_{ii;2}=0$), the rate of change of eigenvalues
and eigenvectors, with respect to various
components of matrix elements of $V$, can be
described as follows:

\begin{eqnarray}
{\partial\lambda_n \over\partial\tau}= -\Omega^2
\lambda_n + 
{\Omega\over h} \sum_{i\leq j} {1\over g_{ij}}
\left[V_{ij;1}(U_{ni} U^{*}_{nj} + U_{nj} U^{*}_{ni})
+ i V_{ij;2}(U_{ni} U^{*}_{nj} - U_{nj}
U^{*}_{ni})\right]
\end{eqnarray}
where
\begin{eqnarray}
{\partial H\over\partial\tau} =-\Omega^2 H
+{V\Omega \over h}
\end{eqnarray}
Further
\begin{eqnarray}
{\partial\lambda_n \over\partial V_{kl;1}}=
{h(\tau)\over \Omega g_{kl}}
\left[ U_{nk} U^{*}_{nl} + U_{nl} U^{*}_{nk} \right] 
\end{eqnarray}

\begin{eqnarray}
{\partial\lambda_n \over\partial V_{kl;2}}=
{h(\tau)\over \Omega g_{kl}}
\left[ U_{nk} U^{*}_{nl} - U_{nl} U^{*}_{nk} \right] \qquad (k \not=
l)
\end{eqnarray}

\begin{eqnarray}
{\partial\lambda_n \over\partial V_{kl;2}}= 0
\qquad ( {\rm if} \quad k =
l)
\end{eqnarray}

and

\begin{eqnarray}
{\partial U_{np} \over\partial V_{kl;1}}=
{-h(\tau)\over \Omega g_{kl}} \sum_{m\not=n}
{1\over {\lambda_m -\lambda_n}}
U_{mp}(U^{*}_{mk}U_{nl} + U^{*}_{ml} U_{nk})
\end{eqnarray}

\begin{eqnarray}
{\partial U_{np} \over\partial V_{kl;2}}=
{-h(\tau)\over \Omega g_{kl}} \sum_{m\not=n}
{1\over {\lambda_m -\lambda_n}}
U_{mp}(U^{*}_{mk}U_{nl} - U^{*}_{ml} U_{nk})
\end{eqnarray}

and
\begin{eqnarray}
\sum_{k\leq l}
\left[{\partial (U_{nk}U^{*}_{nl} + U_{nl}U^{*}_{nk}) \over\partial V_{kl;1}}
+ i {\partial (U_{nk}U^{*}_{nl} -
U_{nl}U^{*}_{nk}) \over\partial V_{kl;2}}
\right] =
-{2 h(\tau)\over \Omega g_{kl}}\sum_{m\not=n}{1
\over{\lambda_n-\lambda_m}}
\end{eqnarray}

	For the real-symmetric case, the corresponding relations 
can be obtained by using $U_{ij}=U_{ij}^*$ (as eigenvector matrix is 
now real-orthogonal) in eqs.(A1-A8) and taking $V_{ij;2}=0$ for all 
values of ${i,j}$ (can also be found in Ref.[9]). 

.
.


\begin{references}


\bibitem{}
B.D.Simons and B.L.Altshuler, Phys. Rev. Lett. 70, 4063 (1993). \\
B.D.Simons, P.A.Lee and B.L.Altshuler, Phys. Rev. Lett. 70, 4122 (1993). 
\bibitem{}
C. Beenakker, Phys. rev. Lett. 70 (1993).
\bibitem{}
K.B.Efetov, Adv. in Phys. 32, 53 (1983).
\bibitem{}
M.L.Mehta {\bf Random Matrice}, (Acedemic, New York, 1991).
\bibitem{}
M.V.Berry, Proc. R. Soc A, 400, (1984).
\bibitem{}
F.Dyson, J. Math. Phys. 3, 1191 (1962).
\bibitem{}
M.Faas, B.D.Simons, X.Zotos and B.L.Altshuler, Phys. Rev. B 48, 5439 (1993).
\bibitem{}
B.Sutherland, J. Math. Phys. 12, 246 (1971); 12, 251, (1971); 
Phys. Rev. A 4 2019 (1971); 5, 1372 (1972).
\bibitem{}
O. Narayan and B.S.Shastry, Phys. Rev. Lett. 71, 2106 (1993).
\bibitem{}
B.S.Shastry, Proceedings of the $16$th Taniguchi 
International Symposium on the Theory of \\ 
Condensed Matter: 
"Correlation Effects in Low Dimensional Electron Systems",\\ 
October 23 to 29, 1993, Shima, Japan (Ed. N.Kawakami and 
A.Okiji, Springer Verlag, 1994). 
\bibitem{}
F.Haake, {\bf Quantum Signatures of Chaos}, (Springer, Berlin, 1991).
\bibitem{}
A.Pandey and P.Shukla, J. Phys. A, (1991).
\bibitem{}
A.Pandey, Chaos, Soliton and Fractals, 5, (1995). 
\bibitem{}
P.Shukla, Submitted to Phys. rev. Lett. (1998).
\bibitem{}
E.Brezin and A.Zee, Nucl. Phys. B402 613 (1993).
\bibitem{}
G.Hackenbroich and H.A.Weidenmuller, Phys. Rev. lett. 74 (1995). 
\bibitem{}
 Present Address: Department of Physics,                       \\
 Indian Institute of Technology,\\ Kharagpur, West Bengal, India  \\
* E-Mail : Shukla@phy.iitkgp.ernet.in
\end{references}
\end{document}